\documentclass{osa-article}

\journal{oe}

\usepackage{array}
\usepackage{enumitem} 
\usepackage{multirow}
\newtheorem{theorem}{Definition}[subsection]
\begin{document}

\title{Topological optimization of hybrid quantum key distribution networks}

\author{Yaxing~Wang, \authormark{1} Qiong~Li, \authormark{1,*} Haokun~Mao, \authormark{1} Qi~Han, \authormark{1} Furong~Huang, \authormark{2} and Hongwei~Xu \authormark{1}}

\address{\authormark{1}School of Computer Science and Technology, Harbin Institute of Technology, Harbin, China}
\address{\authormark{2}School of International Studies, Harbin Institute of Technology, Harbin, China}
\email{\authormark{*}qiongli@hit.edu.cn}




\begin{abstract}
With the growing complexity of quantum key distribution (QKD) network structures, aforehand topology design is of great significance to support a large-number of nodes over a large-spatial area. However, the exclusivity of quantum channels, the limitation of key generation capabilities, the variety of QKD protocols and the necessity of untrusted-relay selection, make the optimal topology design a very complicated task. In this research, a hybrid QKD network is studied for the first time from the perspective of topology, by analyzing the topological differences of various QKD protocols. In addition, to make full use of hybrid networking, an analytical model for optimal topology calculation is proposed, to reach the goal of best secure communication service by optimizing the deployment of various QKD devices and the selection of untrusted-relays under a given cost limit. Plentiful simulation results show that hybrid networking and untrusted-relay selection can bring great performance advantages, and then the universality and effectiveness of the proposed analytical model are verified.
\end{abstract}

\section{Introduction}
Quantum key distribution (QKD) technology, which generates information-theoretic secure (ITS) keys between a long distance communication pair based on the laws of quantum mechanics, has become one of the most promising technologies in quantum communication. However, it is necessary to establish QKD networks \cite{elliott2005current, poppe2008outline, alleaume2009topological, fujiwara2011field, yin2016measurement, liao2017satellite, liao2018satellite, yuan201810, zhang2019continuous} based on multiple QKD devices to provide quantum key service for more users, since a QKD device can only provide quantum keys for a communication pair. In recent years, the number of nodes in experimental QKD networks has expanded from 6 \cite{poppe2008outline, sasaki2011field} to 56 \cite{li2020mathematical}, and the transmission distance has extended from 19.6 \cite{peev2009secoqc} to 2000 kilometers \cite{li2020mathematical}. With the growing complexity of QKD network structures, the design from the perspective of topology is of great significance for quality assurance, cost control, and cycle reduction, etc \cite{dianati2008architecture, diamanti2016practical}.

With the continuous development of QKD technology, more and more types of QKD protocols can be used to construct a QKD network. Different protocols vary greatly in terms of key generation capability, manufacturing cost, fiber dependency and so on. For example, a BB84-QKD device \cite{bennett2014quantum} has a relatively high key generation capability, but the required single-photon detectors are expensive. By contrast, a GG02-QKD device \cite{grosshans2002continuous} does not require single photon detector, which reduces the manufacturing cost, but its key generation capability is lower, especially in the case of long optical fiber distance. Different from the previous two, a TF-QKD device \cite{grasselli2019asymmetric} can achieve a relatively higher key generation rates in asymmetric fiber distances with a relatively low manufacturing cost, by overcoming the point-to-point rate-distance limit \cite{lucamarini2018overcoming}, but it does not work well in the symmetric case. Therefore, facing the diverse fiber distances in complex network topology and the limited cost for network construction, it is worth speculating that a hybrid QKD network composed of multi-type QKD protocols can provide better communication services.

To connect multiple QKD devices, the optical-switch, the quantum-relay, and the trusted-relay are the three commonly used approaches \cite{townsend1997quantum, kumavor2005comparison, ma2006polarization}. Due to the scale limitation of optical-switches\cite{toliver2003experimental} and the technology immaturity of quantum-relays \cite{chen2017experimental, hu2019experimental}, trusted-relay is the most practical approach at present \cite{li2020mathematical}. This approach carries the risk of information leakage at trusted-relays. To reduce the risk, several efforts have been made in recent years \cite{fitzi2007towards, wang2008perfectly, schartner2012quantum}, such as the exclusive-or based key storage \cite{schartner2009overcome} and the game-theory based multi-path transmission \cite{rass2010unified}. However, the premise of the exclusive-or strategy is that the network owner has an efficient node attack detection capability, while the precondition of the game-theory strategy is that the network adversary can only attack limited number of nodes. Therefore, the applicability of these strategies is still limited in the real-life multi-user QKD networks. In order to provide ITS communication service in a real-life hybrid QKD network, the credibility of each trusted-relay must be controlled, which leads to the introduction of credibility control cost. It is generally assumed that all relay nodes are trusted-relays in the previous researches related to trusted-relay based QKD networks \cite{peev2009secoqc, diamanti2016practical}. As a result, the total credibility control cost of a whole network is proportional to the total number of relay nodes. However, for a hybrid QKD network, untrusted-relays, which do not consume credibility control cost, can also be used to connect some types of QKD devices \cite{lo2020scalable}, such as MDI-QKD and TF-QKD. By selecting some cheap untrusted-relays to reduce the total number of trusted relays, the total credibility control cost can be effectively saved. Taking advantage of the saved cost to deploy more QKD devices can certainly generate more quantum keys. Therefore, in contrast to the traditional default scenario where all relay nodes are considered to be trusted-relays, it is worth speculating that efficient selection of untrusted-relays can provide better communication services.

The variety of QKD protocols and the necessity of untrusted-relay selection, coupled with several intrinsic characteristics of QKD devices, such as the exclusivity of quantum channels and the limitation of key generation capabilities \cite{mehic2019novel, lucamarini2018overcoming, li2019improved, mao2019high}, make the optimal design of a hybrid QKD network a very complicated task \cite{maurhart2013new, han2014novel, yang2017qkd,mehic2017implementation,wang2019modeling}. In this research, we study the optimal design of QKD network from the perspective of topology. The main function of topology design is to provide the best communication service within a given cost limit by carefully deploying QKD devices and selecting untrusted-relays. In the literature \cite{alleaume2009topological} and \cite{pederzolli2020optimal}, several cost optimization models for QKD network construction have been designed through the deployment of QKD devices and the construction of new relay nodes. However, it is hard to construct a new relay node according to the optimal working distance calculated by literature \cite{alleaume2009topological} in a real network which is restricted by the physical geography. In addition, MDI-QKD protocol, TF-QKD protocol, and other QKD protocols that rely on two solid fibers have not been studied in these works. Furthermore, all nodes in their QKD networks are considered to be trusted-relays, i.e., the selection of untrusted-relays has not been considered.

\begin{itemize}[leftmargin=1em]
\item In this research, to study the hybrid QKD network, the topological differences between the client-to-client (C2C-QKD) protocol and the client-server-client (CSC-QKD) protocol were analyzed.

\item To make full use of hybrid networking, an analytical model for optimal topology calculation was proposed to calculate the optimal network performance under a given cost limit, and then obtain the optimal deployment of various QKD devices and the selection of untrusted-relays.

\item To verify the universality and effectiveness of this work, a comprehensive simulation based on random topology and real topology was designed. Simulation results demonstrated the advantages of hybrid networking and untrusted-relay selection.
\end{itemize}

The paper is organized as follows: In Section \ref{sec:classification}, the topological differences are presented in detail. Based on the method, an analytical model for optimal topology calculation are proposed in Section \ref{sec:analytical}. In Section \ref{sec:simulation}, the simulations of topology calculation, based on the proposed analytical model, are presented and the results are analyzed. Section \ref{sec:conclusion} presents the concluding remarks.

\section{Topological differences of various QKD protocols}
\label{sec:classification}
\begin{figure}[htbp]
\centering
\includegraphics[width=.85\linewidth]{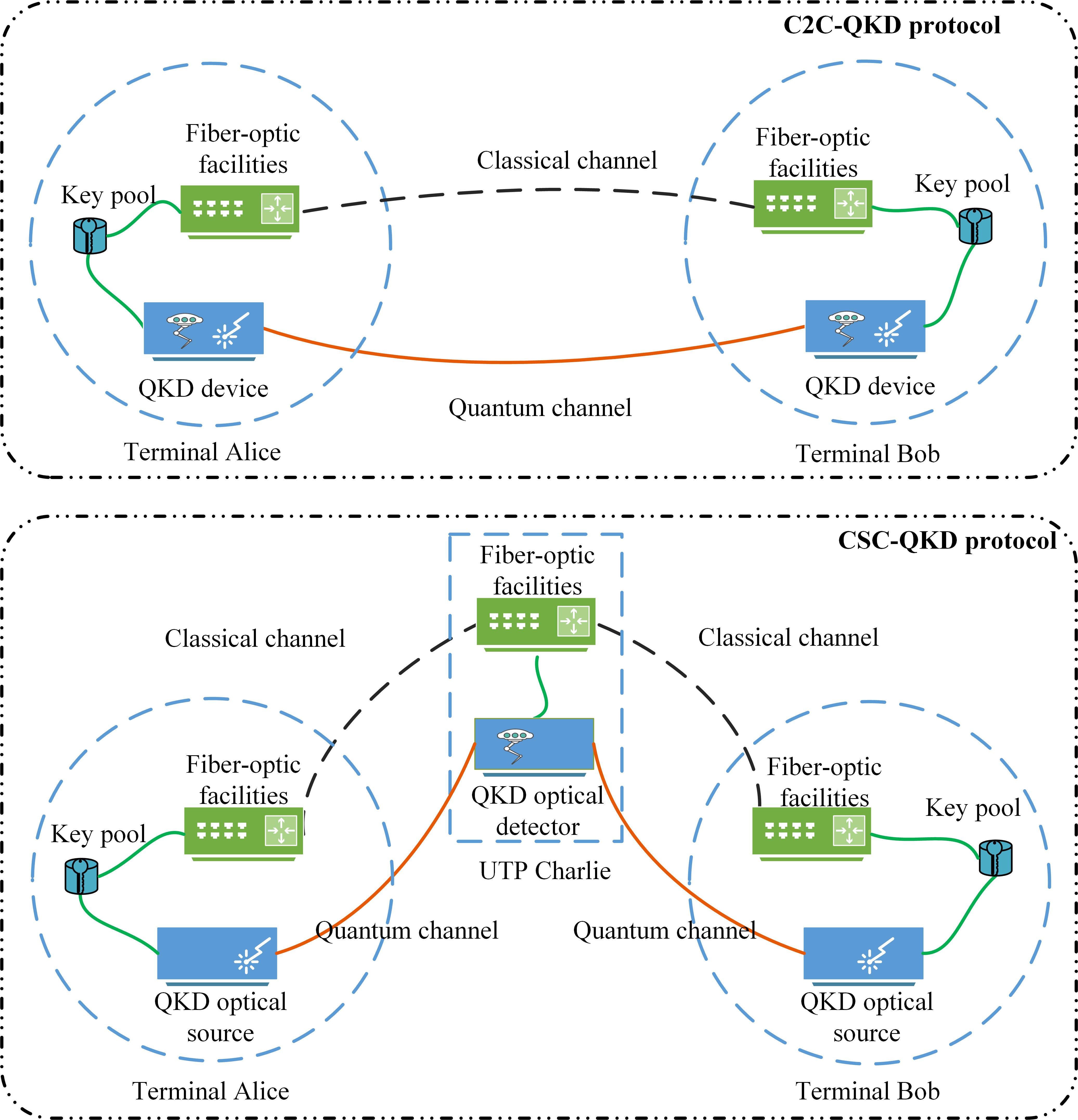}
\caption{The topological differences of various QKD protocols}
\label{fig:1}
\end{figure}

For all types of QKD protocols, quantum state transmission is an important step. Its dependence on quantum channels leads to a typical problem of quantum channel exclusiveness for QKD devices. To solve this problem, researchers have put forward many approaches, such as wavelength-division-multiplexing and orthogonal-frequency-division-multiplexing, to realize the multiplexing of quantum channels on existing optical fibers, so as to provide quantum keys without changing existing communication facilities. As a result, the dependence problem on the existing optical fibers, i.e. fiber dependency, emerges. According to the distinction on fiber dependency, existing various QKD protocols can be divided into C2C-QKD protocol and CSC-QKD protocol, as shown in Fig. \ref{fig:1}.
\vspace{0cm} 
\setlength{\abovecaptionskip}{0.03cm} 
\setlength{\belowcaptionskip}{0cm} 

The C2C-QKD protocol refers to the protocol that only one optical fiber is needed to connect a communication pair, such as BB84-QKD or E91-QKD. In contrast, the CSC-QKD protocol requires the participation of an untrusted third party (UTP), which is actually an untrusted-relay. Meanwhile, both of the two communication parties should connect to the UTP through one optical fiber. MDI-QKD and TF-QKD are the representatives of this kind. For demonstration purposes, we define the two communication parties of a C2C-QKD device as two C2C-clients and label the quantum channel connecting two C2C-clients as a C2C-edge. Similarly, the two communication parties of a CSC-QKD device are defined as two CSC-clients, the UTP is called a CSC-server, and the two quantum channels connecting CSC-clients and CSC-server are labeled as two CSC-edges. In a hybrid QKD network, each node may plays one or more roles as a C2C-client, a CSC-client or a CSC-server, and each edge may contain one or more C2C-edges or CSC-edges.

\section{An analytical model for optimal topology calculation}
\label{sec:analytical}
\subsection{Notations and definitions}
We summarize the notations used throughout the rest of this paper in Table \ref{tab:1} and make the following explanations.
\begin{table}[htbp]
\centering
\caption{Main notations used in the research}
\begin{tabular}{cm{.6\linewidth}c}
\hline
Notation & Explanation & Value\\
\hline
${D^{s,t}}$ & Average communication demand of a communication pair $\left( {s,t} \right)$ &	$R_0^ + $ \\
${\beta ^{s,t}}$ & Ratio of key length to plaintext length in the adopted encryption algorithm & $R_0^ + $ \\
${R_{\left( {u,v} \right)}}$ & Key generation capability of a C2C-QKD device arranged on the C2C-edge $\left( {u,v} \right)$ & $R_0^ + $ \\
${\hat R_{\left( {u,p,v} \right)}}$ & Key generation capability of a CSC-QKD device arranged on the CSC-edge $\left( {u,p,v} \right)$ & $R_0^ + $ \\
$C$ & Available network construction cost & $R_0^ + $ \\
${S_{\left( {u,v} \right)}}$ & Number of C2C-QKD devices arranged on the C2C-edge $\left( {u,v} \right)$ & $N $ \\
${\hat S _{\left( {u,p,v} \right)}}$ & Number of CSC-QKD devices arranged on the CSC-edge $\left( {u,p,v} \right)$ & $N$ \\
$T\left( v \right)$ & Credibility of node $v \in V$ & $\left\{ 0,1 \right\}$ \\
$F_{\left( {u,v} \right)}^{s,t}$ & Flow amount of a communication pair $\left( {s,t} \right)$ passing through the C2C-edge $\left( {u,v} \right)$ & $R_0^ + $ \\
$\hat F _{\left( {u,p,v} \right)}^{s,t}$ & Flow amount of a communication pair $\left( {s,t} \right)$ passing through the CSC-edge $\left( {u,p,v} \right)$ & $R_0^ + $ \\
$B$ & Global satisfaction degree of all communication demands & $R_0^ + $ \\
\hline
\end{tabular}
\label{tab:1}
\end{table}

To expand the discussion more smoothly, it is necessary to explain in detail the definition and meaning of SoD, G-SoD and MG-SoD, the three main concepts used for topology evaluation, as shown below.

\begin{theorem}[SoD]
\label{SoD}
For a communication pair $\left( {s,t} \right)$, its satisfaction degree of communication demand (SoD), ${B^{s,t}}$, is the ratio of its total key consumption, ${A^{s,t}}$, to its total key demand \cite{li2020mathematical}, i.e.,
\begin{equation}
\label{eq:new 1}
{B^{s,t}} = \frac{{{A^{s,t}}}}{{{D^{s,t}} \cdot {\beta ^{s,t}}}},
\end{equation}
where ${A^{s,t}}$ can be calculated by the total difference between the flows into and out of the source node $s$, i.e.,
\begin{equation}
\label{eq:new 2}
{A^{s,t}} = \sum\limits_{v \in V} {\left[ {F_{\left( {s,v} \right)}^{s,t} - F_{\left( {v,s} \right)}^{s,t}} \right]} + \sum\limits_{p \in V,v \in V} {\left[ {\hat F _{\left( {s,p,v} \right)}^{s,t} - \hat F _{\left( {v,p,s} \right)}^{s,t}} \right]}.
\end{equation}
\end{theorem}

\begin{theorem}[G-SoD]
\label{G-SoD}
For a QKD network with a given topology $G = \left( {V,E} \right)$, its global satisfaction degree of all communication demands (G-SoD), $B$, is defined as the worst performance of all communication pairs \cite{li2020mathematical}. Thus, $B$ can be calculated by the minimum value of all SoDs, i.e.,
\begin{equation}
\label{eq:new 3}
B = \mathop {\min }\limits_{s \in V,t \in V} {B^{s,t}} = \mathop {\min }\limits_{s \in V,t \in V} \frac{{{A^{s,t}}}}{{{D^{s,t}} \cdot {\beta ^{s,t}}}}.
\end{equation}
\end{theorem}

\begin{theorem}[MG-SoD]
\label{MG-SoD}
For a QKD network topology $G = \left( {V,E} \right)$, its optimal performance is defined as the maximum of all possible G-SoDs (MG-SoD), each of which is defined as the worst performance of all communication pairs over a possible flow assignment \cite{li2020mathematical}.
\end{theorem}

\subsection{The proposed analytical model}
To make full use of hybrid networking, an analytical model, which is essentially an optimization formulation, is designed to solve the problem of the optimal deployment of various QKD devices and the selection of untrusted-relays for a hybrid QKD network.

The input items of this formulation mainly include the existing classical network, the communication demand, the key generation capability and the limited construction cost. (i) The existing classical network is represented as a directed graph $G = \left( {V,E} \right)$. Then, the set of all optional C2C-edges is $E$ and the set of all optional CSC-edges is $\hat E = \left\{ {\left( {u,p,v} \right)|\left( {u,p} \right) \in E,\left( {p,v} \right) \in E} \right\}$. (ii) The communication demand is defined as ${D^{s,t}}$ and the key consumption ratio as ${\beta ^{s,t}}$. Therefore, the requested key demand is ${D^{s,t}} \cdot {\beta ^{s,t}}$ for each communication pair $\left( {s,t} \right)\left( {s \in V,t \in V} \right)$. In particular, ${\beta ^{s,t}} = 1$ indicates that the one-time-pad algorithm is adopted to achieve ITS communication and ${\beta ^{s,t}}= 0$ indicates that the adopted encryption algorithm does not require quantum keys. (iii) The key generation capability for each C2C-QKD device arranged on the C2C-edge $\left( {u,v} \right) \in E$ is denoted as ${R_{\left( {u,v} \right)}}$, and the key generation capability for each CSC-QKD device arranged on the CSC-edge $\left( {u,p,v} \right) \in \hat E $ is denoted as ${\hat R _{\left( {u,p,v} \right)}}$. (iv) The total cost for QKD network construction is limited to $C$, which mainly includes the manufacturing cost of QKD devices and the credibility control cost of trusted-relays.

The decision variables of this formulation mainly contain three types: integer variable, boolean variable and continuous variable. (i) The integer variable ${S_{\left( {u,v} \right)}}$ represents the number of C2C-QKD devices arranged on the C2C-edge $\left( {u,v} \right)$, and ${\hat S _{\left( {u,p,v} \right)}}$ represents the number of CSC-QKD devices arranged on the CSC-edge $\left( {u,p,v} \right)$. (ii) The boolean variable $T\left( v \right)$ indicates whether node $v$ is a trusted-relay and whether credibility control is required for this node. (iii) The continuous variable $F_{\left( {u,v} \right)}^{s,t}$ represents the flow amount of the communication pair $\left( {s,t} \right)$ \cite{schrijver2002history, han2014maximum, li2020mathematical} passing through the C2C-edge $\left( {u,v} \right)$, $\hat F _{\left( {u,p,v} \right)}^{s,t}$ represents the flow amount of the communication pair $\left( {s,t} \right)$ passing through the CSC-edge $\left( {u,p,v} \right)$, and B represents the quality of a QKD network topology, which can be measured by the G-SoD \cite{li2020mathematical}.

The complete formulation is then given by,

{\setlength{\abovedisplayskip}{0pt}
\setlength{\belowdisplayskip}{0pt}
Maximize:
\begin{equation}
\label{eq:1}
B.
\end{equation}

Subject to:
\begin{equation}
\label{eq:2}
\begin{aligned}
\forall \left( {u,v} \right) \in E,0 \le \sum\limits_{s \in V,t \in V} {\left[ {F_{\left( {u,v} \right)}^{s,t} + F_{\left( {v,u} \right)}^{s,t}} \right]} \le {S_{\left( {u,v} \right)}}{R_{\left( {u,v} \right)}} + {S_{\left( {v,u} \right)}}{R_{\left( {v,u} \right)}},
\end{aligned}
\end{equation}

\begin{equation}
\label{eq:3}
\begin{aligned}
\forall \left( {u,p,v} \right) \in \hat E, 0 \le \sum\limits_{s \in V,t \in V} {\left[ {\hat F _{\left( {u,p,v} \right)}^{s,t} + \hat F _{\left( {v,p,u} \right)}^{s,t}} \right]} \le {\hat S _{\left( {u,p,v} \right)}}{\hat R _{\left( {u,p,v} \right)}} + {\hat S _{\left( {v,p,u} \right)}}{\hat R _{\left( {v,p,u} \right)}},
\end{aligned}
\end{equation}

\begin{equation}
\label{eq:4}
\begin{aligned}
\forall s \in V,\forall t \in V,\forall u \in \left( {V - \left\{ {s,t} \right\}} \right),\sum\limits_{v \in V} {\left[ {F_{\left( {u,v} \right)}^{s,t} - F_{\left( {v,u} \right)}^{s,t}} \right]} + \sum\limits_{p \in V,v \in V} {\left[ {\overline F _{\left( {u,p,v} \right)}^{s,t} - \overline F _{\left( {v,p,u} \right)}^{s,t}} \right]} = 0,
\end{aligned}
\end{equation}

\begin{equation}
\label{eq:5}
\begin{aligned}
\forall s \in V,\forall t \in V, \sum\limits_{v \in V} {\left[ {F_{\left( {s,v} \right)}^{s,t} - F_{\left( {v,s} \right)}^{s,t}} \right]} + \sum\limits_{p \in V,v \in V} {\left[ {\hat F _{\left( {s,p,v} \right)}^{s,t} - \hat F _{\left( {v,p,s} \right)}^{s,t}} \right]} = {A^{s,t}},
\end{aligned}
\end{equation}

\begin{equation}
\label{eq:6}
\begin{aligned}
\forall s \in V,\forall t \in V, \sum\limits_{v \in V} {\left[ {F_{\left( {t,v} \right)}^{s,t} - F_{\left( {v,t} \right)}^{s,t}} \right]} + \sum\limits_{p \in V,v \in V} {\left[ {\hat F _{\left( {t,p,v} \right)}^{s,t} - \hat F _{\left( {v,p,t} \right)}^{s,t}} \right]} = - {A^{s,t}},
\end{aligned}
\end{equation}

\begin{equation}
\label{eq:7}
\forall s \in V,\forall t \in V,{A^{s,t}} - B \cdot {D^{s,t}} \cdot {\beta ^{s,t}} \ge 0,
\end{equation}

\begin{equation}
\label{eq:8}
\sum\limits_{\left( {u,v} \right) \in E} {{S_{\left( {u,v} \right)}}} + {q_1}\sum\limits_{\left( {u,p,v} \right) \in \hat E } {{{\hat S }_{\left( {u,p,v} \right)}}} + {q_2}\sum\limits_{v \in V} {T\left( v \right)} - C \le 0,
\end{equation}

\begin{equation}
\label{eq:9}
\forall v \in V, T\left( v \right) =
\begin{cases}
0 & \text{if } I\left( v \right) = 0,\\
1 & \text{if } I\left( v \right) \ne 0,
\end{cases}
\end{equation}
where
\begin{equation}
\begin{aligned}
\label{eq:10}
I\left( v \right) = \sum\limits_{\left( {u,v} \right) \in E} {{S_{\left( {u,v} \right)}}} + \sum\limits_{\left( {v,u} \right) \in E} {{S_{\left( {v,u} \right)}}} + \sum\limits_{\left( {u,p,v} \right) \in \hat E } {{{\hat S }_{\left( {u,p,v} \right)}}} + \sum\limits_{\left( {v,p,u} \right) \in \hat E } {{{\hat S }_{\left( {v,p,u} \right)}}}.
\end{aligned}
\end{equation}}

The objective function is expressed in Eq. (\ref{eq:1}), which maximizes the G-SoD. Constraints Eqs. (\ref{eq:2}, \ref{eq:3}) state the capacity constraint condition, i.e., the total flow amount through an edge must be lower than the total key generation capability on this edge. Constraint Eq. (\ref{eq:4}) expresses the flow conservation condition, i.e., the total flow amount that enters a non-source-sink node must be equal to the total flow amount that exits this node. Constraints Eqs. (\ref{eq:5}-\ref{eq:7}) prove that sufficient key materials from the source node to the sink one is available to satisfy the key demand between the two nodes. Constraint Eq. (\ref{eq:8}) requires the actual cost must be no higher than the limited cost. Here, we use the price of a C2C-QKD device as the basis of cost calculation. Besides, the price of a CSC-QKD device is assumed to be ${q_1}$ times the base value, and the credibility control cost of a trusted-relay be ${q_2}$ times the base value. Constraint Eq. (\ref{eq:9}) carries out the selection of untrusted-relays, which have neither been C2C-client nor CSC-Client.

In this formulation, all other constraints are linear functions, except the Eq. (\ref{eq:9}) which is a piecewise function. To facilitate the solution, we transform the Eq. (\ref{eq:9}) into a linear expression by means of approximation, making this formulation a standard mixed-integer linear programming (MILP) model. More specifically, the Eq. (\ref{eq:9}) is rewritten as Eq. (\ref{eq:11}) by introducing additional boolean decision variable $T'\left( v \right)$ and continuous decision variable $T''\left( v \right)$, where $M$ is an arbitrarily large real number.

\begin{equation}
\label{eq:11}
\left\{ \begin{array}{l}
\forall v \in V,T'\left( v \right) + T''\left( v \right) = 1,\\
\forall v \in V,I\left( v \right) - M \cdot T\left( v \right) \le 0,\\
\forall v \in V,I\left( v \right) + T'\left( v \right) - T''\left( v \right) \ge 0.
\end{array} \right.
\end{equation}
The decision variables of this standard MILP formulation are listed below:
\begin{equation}
\label{eq:14}
\left\{ \begin{array}{l}
B \in R_0^ +,\\
\forall \left( {u,v} \right) \in E,\forall s \in V,\forall t \in V,F_{\left( {u,v} \right)}^{s,t} \in R_0^ +,\\
\forall \left( {u,p,v} \right) \in \hat E,\forall s \in V,\forall t \in V, \hat F _{\left( {u,p,v} \right)}^{s,t} \in R_0^ +,\\
\forall \left( {u,v} \right) \in E,S\left( {u,v} \right) \in N,\\
\forall \left( {u,p,v} \right) \in \hat E,\hat S \left( {u,p,v} \right) \in N,\\
\forall v \in V,T\left( v \right) \in \left\{ {0,1} \right\},\\
\forall v \in V,T'\left( v \right) \in \left\{ {0,1} \right\},\\
\forall v \in V,T''\left( v \right) \in R_0^ +.
\end{array} \right.
\end{equation}

As a typical MILP model, this formulation can be solved by a mature linear programming solver, Gurobi \cite{li2018combinatorial, helm2018extension}. After optimization, the proposed model is convenient to retrieve the MG-SoD (i.e., the objective function value), the number of QKD devices to be installed on each edge (i.e., ${S_{\left( {u,v} \right)}}$ and ${\hat S _{\left( {u,p,v} \right)}}$), the nodes to be selected as untrusted-relays (i.e., $T\left( v \right)$), and the paths to be selected for each communication pair to route the encrypted messages (i.e., $F_{\left( {u,v} \right)}^{s,t}$ and $\hat F _{\left( {u,p,v} \right)}^{s,t}$).

\section{Simulation results and analysis}
\label{sec:simulation}
\subsection{Optimal topology calculation of several random graphs}
We evaluate the calculation results of our proposed analytical model on 13 random graphs, which are generated based on the ER model \cite{garlaschelli2009weighted, drobyshevskiy2019random}. These random graphs are set as follows: each family consists of 10 instances with a fixed number of nodes, an average nodal degree of 3, a set of edge lengths uniformly distributed at $\left[ {10,500} \right]$ kilometers, and a uniform traffic matrix established at $\left[ {100,500} \right]$ kbps between one fifth of nodes allowing the remaining four-fifths the opportunity to act as untrusted-relays. The simulation platform adopted in this research is detailed in Table \ref{tab:2}.

\begin{table}[htbp]
\centering
\caption{The simulation platform}
\begin{tabular}{ccc}
\hline
 & Parameter & Value\\
\hline
\multirow{2}*{CPU} & Version & Intel(R) Core(TM) i9-9820X \\
		~ & Performance & 10-Cores@3.3GHz \\
\hline
\multirow{3}*{Gurobi} & TimeLimit & 7200s \\
		~ & SolutionLimit & 200 \\
		~ & MIPGap & 0.01 \\
\hline
\end{tabular}
\label{tab:2}
\end{table}

In addition, we adopt the decoy-QKD protocol \cite{wang2019modeling} to prepare C2C-QKD devices, and the SNS-TF-QKD \cite{grasselli2019asymmetric} protocol to prepare CSC-QKD devices. To simplify the analysis, we assume that the parameters of all C2C-QKD devices are the same, and the parameters of all CSC-QKD devices are also the same. It should be noted that the proposed analytical model supports the configuration of different parameter setting for each QKD device. The main cost of a decoy-QKD device mainly comes from two single-photon detectors, similarly, the main cost of a SNS-TF-QKD device also mainly comes from the same detectors. Therefore, ${q_1}$ is assumed to be 1, by taking the price of a decoy-QKD device as the basis of cost calculation. Considering the complexity of credibility control, which is usually realized by physical protection \cite{zhang2018large, techateerawat2011network, evans2019demonstration, zhang2012real, mailloux2015modeling}, ${q_2}$ is assumed to be 100. In addition, the overall cost for network construction is limited to 10000.

To demonstrate the advantages of hybrid networking and untrusted-relay selection clearly, the average values of all MG-SoDs obtained from 13 families of random graphs are standardized. Specifically, the simulation result in the case of hybrid networking and untrusted-relay selection is normalized as 100\%, and the remaining results are normalized as the ratio to the above values, for each graph size.

\begin{figure}[htbp]
\centering
\includegraphics[width=.7\linewidth]{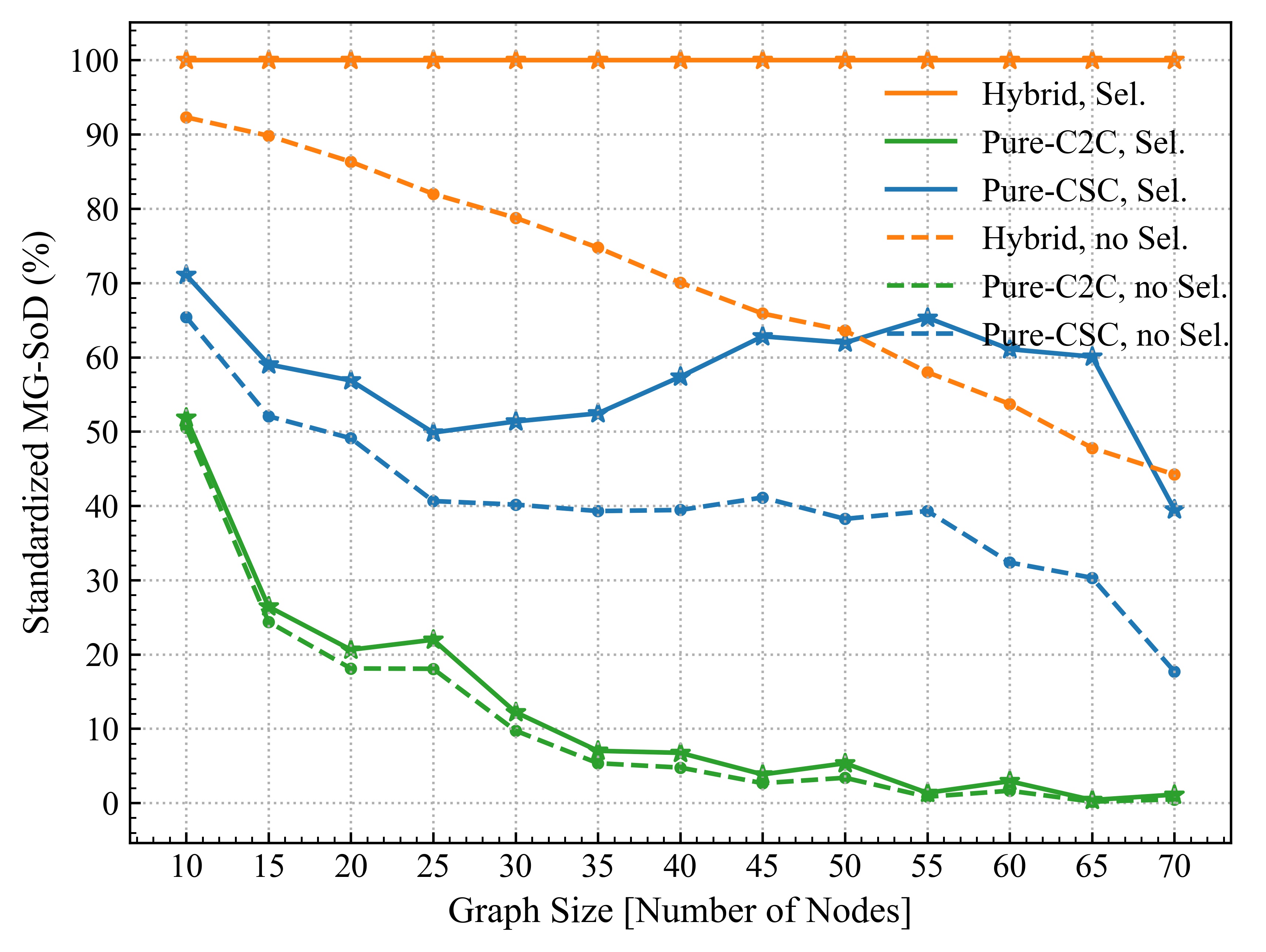}
\captionsetup{width=.9\linewidth}
\caption{Standardized MG-SoDs. The orange curves show the values for hybrid networks which contains both CSC-QKD and C2C-QKD devices; the blue curves denote the values for pure-CSC networks consisting of CSC-QKD devices only; the green curves denote the values for pure-C2C networks consisting of C2C-QKD devices only. The solid lines and dotted lines represent the cases with or without untrusted-relay selection respectively.}
\label{fig:2}
\end{figure}

In Fig. \ref{fig:2}, we report the standardized MG-SoDs of hybrid networks, pure-C2C networks and pure-CSC networks versus the graph size with or without untrusted-relay selection, respectively. As can be seen from the figure, higher MG-SoDs can be achieved when using multi-type QKD protocols: the value of hybrid network curves, which are orange, is always higher than that of green and blue curves. Higher MG-SoDs can be achieved by carefully selecting untrusted-relays: the value of the solid lines is always higher than the dotted lines. In addition, as the number of nodes increases, the performance of pure-C2C networks drops sharply, while such performance drop of pure-CSC networks is not always obvious. Furthermore, without untrusted-relay selection, the performances of all three types of QKD networks decrease monotonically with the increase of graph size.

\begin{figure}[htbp]
\centering
\includegraphics[width=.7\linewidth]{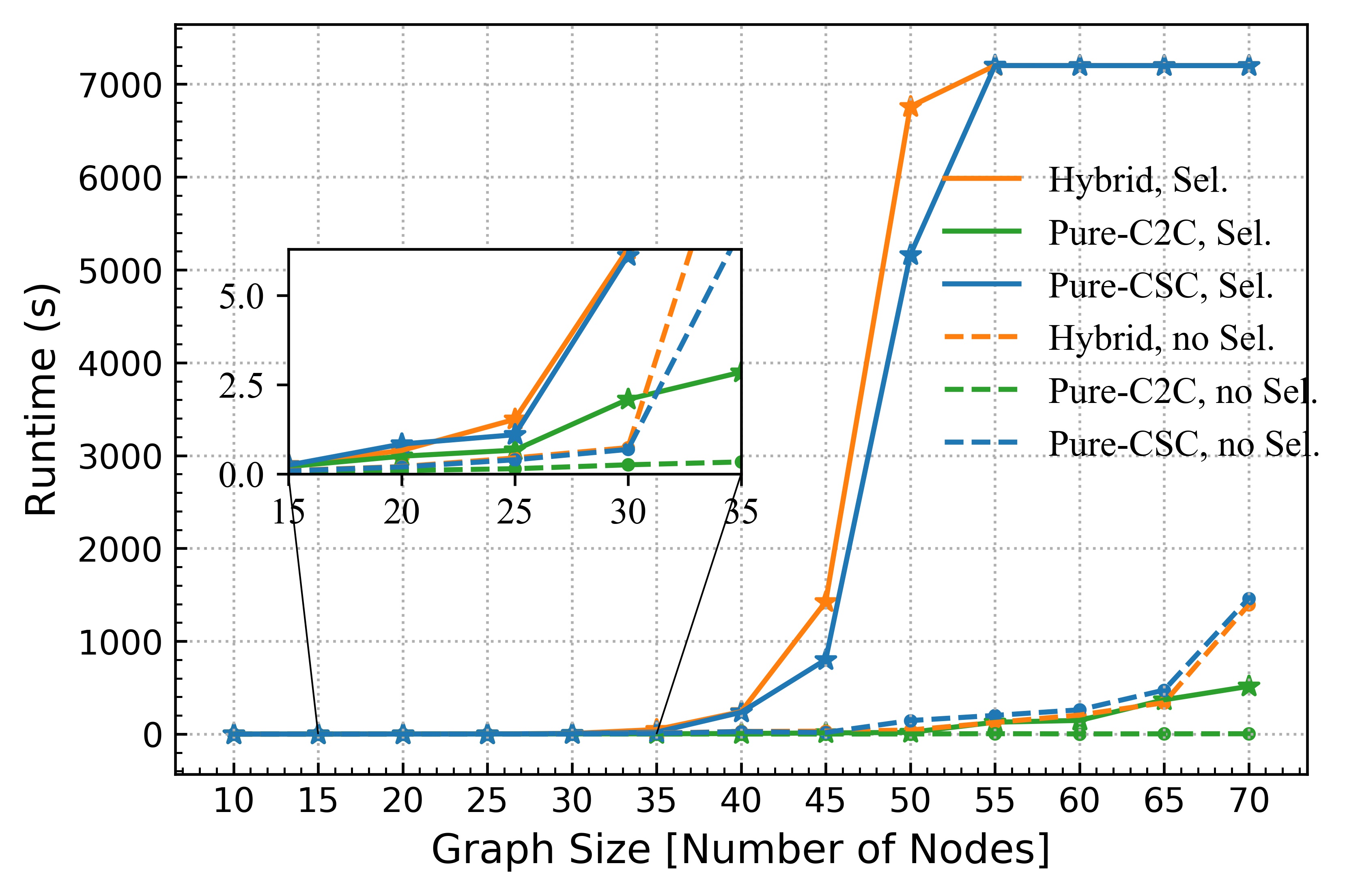}
\captionsetup{width=.9\linewidth}
\caption{Runtimes of different networking schemes. The orange curves show the values for hybrid networks; the blue curves denote the values for pure-CSC networks; the green curves denote the values for pure-C2C networks. The solid lines and dotted lines represent the cases with or without untrusted-relay selection respectively.}
\label{fig:3}
\end{figure}

To show the solving difficulty of our proposed analytical model more intuitively, we report the runtime results of different networking schemes in Fig. \ref{fig:3}. As can be seen from the figure, the addition of hybrid networking and untrusted-relay selection makes the solving more difficult: the value of hybrid network curves, which are orange, is always higher than that of green and blue curves, and the value of the solid lines is always higher than the dotted lines. More obviously, as the number of nodes increases, the runtime of hybrid networks with untrusted-relay selection is always the longest, followed by the pure-CSC networks with untrusted-relay selection. By contrast, the runtimes of the remaining four schemes are much shorter, because they have fewer decision variables and fewer constraints. In addition, the optimal solution cannot be obtained within 7200 seconds, when the number of nodes reaches 55. It is worth to mention that, it is essential to design corresponding heuristic algorithm \cite{holmberg2000lagrangian,yaghini2012simplex} to accelerate the calculation of optimal solution for large-scale network construction, which is also the research focus of our future work.

\subsection{Optimal topology calculation of a real NSFNET topology}
\begin{figure}[htbp]
\centering
\includegraphics[width=.7\linewidth]{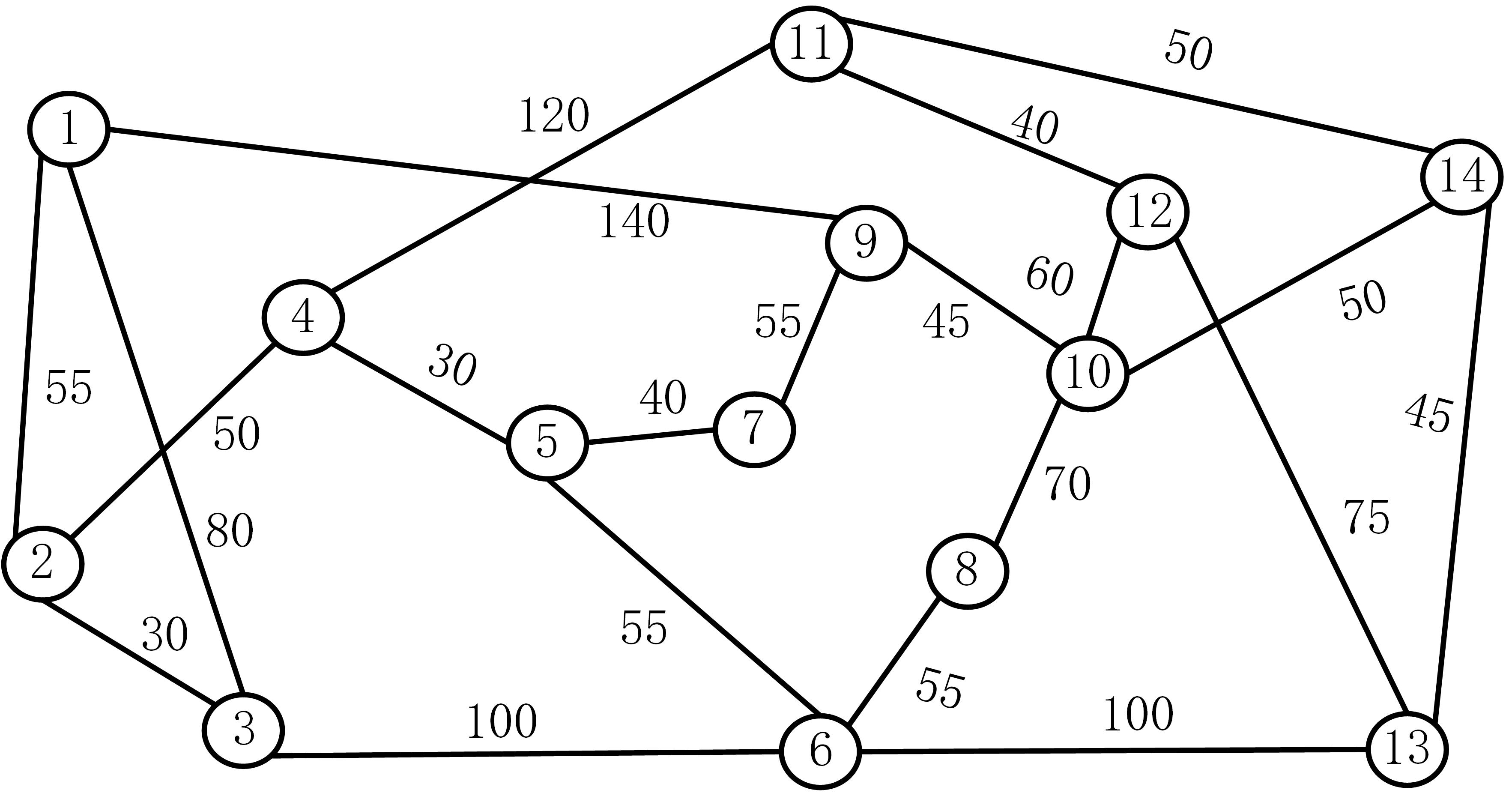}
\caption{Topology of NSFNET network.}
\label{fig:4}
\end{figure}

Although the above results are sufficient to prove the universality of this study, another simulation based on a real NSFNET topology (14 nodes and 21 edges) \cite{li2020mathematical}, as shown in Fig. \ref{fig:4}, is conducted to further verify the effectiveness of the proposed analytical model.

Under the same assumptions, the corresponding optimal design is shown in Table \ref{tab:3}, where Hyd. SoD, C2C. SoD, and CSC. SoD are used to indicate the MG-SoDs of hybrid network, pure-C2C network and pure-CSC network, respectively. From the data in the table, it can be concluded that both the hybrid networking and the untrusted-relay selection play a critical role in optimal topology design; therefore, the effectiveness of this study is verified.

\begin{table}[htbp]
\centering
\caption{Optimal design under NSFNET topology}
\begin{tabular}{cccc}
\hline
Selection & Hyd. SoD & C2C. SoD & CSC. SoD \\
\hline
No & 828.12 & 740.95 & 335.93\\
Yes & 915.65 & 785.09 & 383.37\\
\hline
\end{tabular}
\label{tab:3}
\end{table}

Furthermore, it should be pointed out that, although the above simulation only gives the results of a set of specific input parameters due to the length limit, the conclusions are also established for various input parameters.

\section{Conclusion}
\label{sec:conclusion}
In conclusion, we have proposed an analytical model to find the optimal deployment of various QKD devices and the selection of untrusted-relays under a given cost limit, and then to create a high-quality hybrid QKD network with the existing classical network. Through simulation, the universality and effectiveness of this study have been verified. In the process of research, the proposed model is calculated by using existing computing tools, Gurobi. However, in the face of large-scale QKD networks, specific heuristic algorithms should be designed to accelerate the calculation of the proposed model, which is also the research focus of our future work. In addition, to further improve the topology performance, we will study the physical protection of trusted-relays thoroughly based on existing works, to quantify and reduce the physical protection cost in our future work. We anticipate that the proposed model could contribute a significance for researchers to accelerate the construction process of QKD networks.

\section*{Funding}
Space Science and Technology Advance Research Joint Funds (6141B06110105); National Natural Science Foundation of China (NSFC) (61301099).

\section*{Disclosures}
The authors declare no conflicts of interest.

\bibliography{reference}


\end{document}